\documentclass[twocolumn,english,showpacs]{revtex4-2}

\usepackage{graphicx}%
\usepackage{multirow}%
\usepackage{array}
\usepackage{amsmath,amssymb,amsfonts}%
\usepackage{amsthm}%
\usepackage{mathrsfs}%
\usepackage[title]{appendix}%
\usepackage{xcolor}%
\usepackage{textcomp}%
\usepackage{booktabs}%
\usepackage{algorithm}%
\usepackage{algorithmicx}%
\usepackage{algpseudocode}%
\usepackage{listings}%
\usepackage[bookmarks=false]{hyperref}
\usepackage{booktabs}
\usepackage{siunitx}
\usepackage{makecell}
\usepackage{threeparttable}
\newcommand{\ket}[1]{\vert{#1}\rangle}
\newcommand{\bra}[1]{\langle{#1}|}

\newcommand{\Ba}{$^{138}$Ba$^+$~}

\newcommand{\0}{\ket{^2S_{1/2}, m_J =-\frac{1}{2}}}

\newcommand{\X}{\ket{^2S_{1/2}, m_J =+\frac{1}{2}}}
\newcommand{\Xp}{\ket{^2D_{5/2}, m_J =+\frac{3}{2}}}
\newcommand{\e}{\ket{^2P_{1/2}, m_J =+\frac{1}{2}}}
\newcommand{\states}{\ket{^2D_{5/2}, m_J = \{ -\frac{1}{2},-\frac{3}{2}, +\frac{1}{2} \} }}
\newcommand{\DQC}{Duke Quantum Center, Department of Electrical and Computer Engineering and Department of Physics, Duke University, Durham, NC 27708}

\begin{document}

\title{Photonic Networking of Quantum Memories in High-Dimensions}

\author{Mikhail Shalaev}
\email{Corresponding author: mikhail.shalaev@duke.edu}
\author{Sagnik Saha}
\author{George Toh}
\author{Isabella Goetting}
\author{Ashish Kalakuntla}
\author{Harriet Bufan Shi}
\author{Jameson O'Reilly}
\email{Present Address: Department of Physics, University of Oregon, Eugene, OR 97331}
\author{Yichao Yu}
\author{Christopher Monroe}
\affiliation{\DQC}

\date{\today}

\begin{abstract} 
Quantum networking enables the exchange of quantum information between physically separated quantum systems \cite{review_Azuma2023}, which has applications ranging from quantum computing to unconditionally secure communication.
Such quantum information is generally represented by two-level quantum systems or qubits.
Here, we demonstrate a quantum network of high-dimensional (HD) quantum memories or ``qudits" stored in individual atoms. 
The interference and detection of HD time-bin encoded single photons emitted from atomic qudit memories heralds maximally-entangled Bell states across pairs of atomic qudit levels.
This approach expands the quantum information capacity of a quantum network while improving the entanglement success fraction beyond the standard 50\% limit of qubit-based measurement protocols \cite{Calsamiglia2001}.
\end{abstract}

\maketitle

Scalable quantum computing will likely require optical photonic interconnects for modular and replicable architectures, regardless of the underlying physical platform.
In such architectures, quantum memories store and process quantum information while single photons distribute quantum information across the network~\cite{Monroe2013, Covey2023}.
While most quantum information processing rely on two-level systems or qubits, many quantum platforms naturally possess multi-level structures that can serve as HD quantum memories. 
Qudits are quantum memories with $d$ distinct levels, and $n$ qudits offer access to a $d^n$-dimensional Hilbert space.
Qudits promise more efficient resource utilization, enable novel algorithms for quantum computing and simulations~\cite{Qudit_logic_Ringbauer2022, Qudit_logic_Hrmo2023, Senko2020, Senko2023, Meth2025}, and support new protocols for quantum networking~\cite{HD_photons_Erhard2020, HD_ent_rate_Yamazaki2024} and quantum key distribution~\cite{QKD_Cerf2002, QKD_Bacco2024}.

Many quantum platforms support qudit memories. 
Superconducting Josephson junctions have natural qudits based on the anharmonicity of their oscillators~\cite{Morvan2021, Blok2021}, although it remains a great challenge to couple them to optical photons with high fidelity \cite{Mirhosseini2020, Krastonov2021}.  
Entanglement between atomic ensemble qudits has been explored~\cite{hyper_ent_mems_Zhang2016}, although such qudits cannot readily be directly manipulated or entangled with other qudit memories.
HD single photons have been realized using various degrees of freedom such as time, frequency, and spatial mode~\cite{HD_photons_Erhard2020}, and two photons have been entangled in a $(100 \times 100)$-dimensional Hilbert space~\cite{HD_photons_Krenn2014}.
Atoms offer access to multiple long-lived metastable states suitable for qudit encoding~\cite{Saffman2008, Covey2024, Deutsch2021, Qudit_logic_Ringbauer2022, Qudit_logic_Hrmo2023, Senko2020, Senko2023, Meth2025} and are a natural interface to pure single photons~\cite{Crocker2019} enabling quantum networking~\cite{Stephenson2020, high_F_Saha2024, Ritter2012, Van_Leent_atom_ent_2022}. Moreover, atoms are perfectly replicable quantum memories that can exhibit excellent coherence~\cite{coh_time_Wang2021}, nearly perfect state preparation and measurement~\cite{state_prep_Sotirova2024}, and high-performance local entangling gate operations~\cite{industry_Loschnauer2024, finkelstein2024}.

In this paper, we generate HD time-bin-encoded single photons entangled with single atomic qudit memories, up to $d=4$ dimensions.
With two such physically separated atom-photon qudit systems, we probabilistically generate Bell-like remote entanglement between the two atomic qudit memories by interfering and detecting the two HD photons. We demonstrate qudit entanglement fidelities up to $0.987(13)$. 
Finally, we directly verify the expected photonic qudit heralding success fraction $\mathscr{F}=1-1/d$ , exceeding the standard Bell-state generation limit of $1/2$ for qubits ($d=2$).

\section*{Qudit memory-photon interface}

\begin{figure*}[htb!]%
\centering
\includegraphics[width=0.9\textwidth]{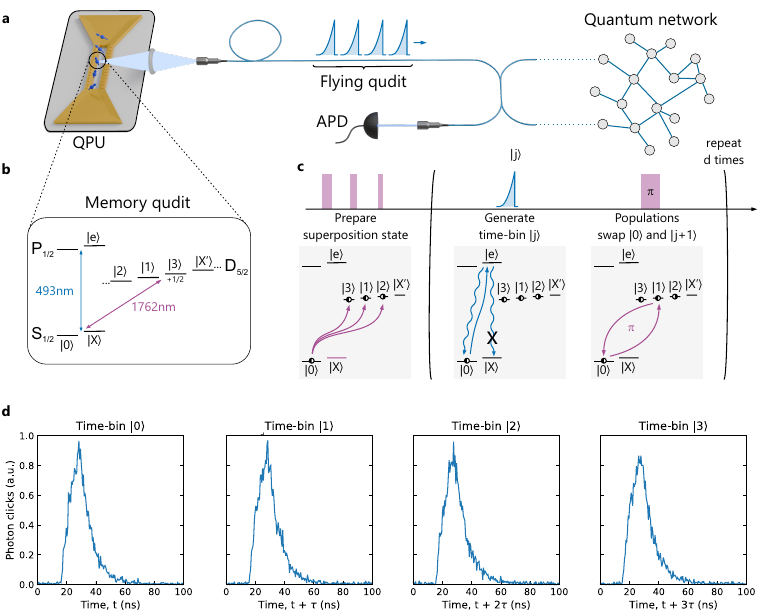}
\caption{
\textbf{Single atom-photon interface. (a)} A qudit-based quantum processing unit (QPU) containing quantum memories stored in multiple electronic states of trapped atomic ions.
The populations stored in these levels are entangled with high-dimensional (HD) time-bin-encoded single photons.
These photons serve as flying qudits for quantum networking applications.
\textbf{(b)} Relevant energy levels of a \Ba atomic ion (nuclear spin $I=0$).
The qudit basis states $\ket{0}$, $\ket{1}$, $\ket{2}$, and $\ket{3}$ are encoded in the $^2S_{1/2}$ and $^2D_{5/2}$ manifolds and manipulated using a narrow-band $1762$~nm laser. 
Lasers operating at $493$~nm and $650$~nm (for repumping from the $^2 D_{3/2}$ state not shown) are used for Doppler cooling, state initialization, and state readout. 
\textbf{(c)} Atom-photon entanglement protocol. The atomic qudit is first prepared in a superposition of all its states with coherent 1762 nm pulses. An ultrafast pulsed laser at $493$~nm excites the atomic ion from $\ket{0}$ to the short-lived excited state $\ket{e}$, resulting in single-photon emission for the first time-bin correlated with state $\ket{0}$ (decays to $\ket{X}$ are vetoed through polarization selection and erasure correction as described in the text.) Next, the qudit populations are swapped with additional 1762 nm pulses, and the excitation step is repeated to generate the other time bins.
\textbf{(d)} Distribution of arrival times of 6106 photon events appearing in four time-bins separated by $\tau = 5680$~ns, measured with an avalanche photodetector (APD).
Each peak corresponds to a distinct time-bin with a 7.86 ns exponential decay, broadened by electronic and detector response delays.
}
\label{fig: photon bins}
\end{figure*}

We begin by generating single photons encoded in up to four distinguishable time windows, or time-bins, emitted from a single atom (Fig.~\ref{fig: photon bins}a).
Specific atomic levels are coherently correlated with the corresponding time-bins of a single photon, enabling entanglement between atom and photon.
We encode the qudit state in up to four ground and metastable levels of a single \Ba atomic ion defined by $\ket{0} \equiv \0$ and $\ket{ \{1, 2, 3\} } \equiv \states$, as shown in Fig.~\ref{fig: photon bins}b.
We also include ground and metastable error states $\ket{X} \equiv \X$ and $\ket{X'} \equiv \Xp$, which are used to veto infrequent erasure errors during photon generation, described below \cite{high_F_Saha2024}. 

Fig.~\ref{fig: photon bins}c shows the experimental protocol and pulse sequence for generating entanglement between the atom’s internal levels and a single photon's time-bin degree of freedom.
The protocol begins by Doppler cooling the atomic ion using $493$~nm and $650$~nm light.
The atom is optically pumped to the $\ket{0}$ state and initialized in an equal superposition state $\ket{\psi_0} = \frac{1}{\sqrt{d}} \sum_{j=0}^{d-1} \ket{j}$ with appropriate pulses of light at 1762 nm.
Next, a 3~ps $\sigma^+$-polarized 493 nm optical pulse excites population from $\ket{0}$ to the state $\ket{e} \equiv \e$ which then spontaneously decays back to $\ket{0}$, emitting an exponentially-decaying wave packet with characteristic lifetime 7.86 ns thus generating the first time-bin of a single photon.
Additional time-bins are generated by subsequently swapping the population between $\ket{0}$ and $\ket{j}$ states, followed by an excitation pulse as before~\cite{barrett_kok}.
The time-bin separation is made synchronous with the atomic ion's motional modes to avoid entanglement with atomic motion, which could otherwise degrade photon coherence~\cite{high_F_Saha2024, motion_theory_Yu2025, motion_theory_Kikura2025, motion_theory_Apolín2025}.  
Fig.~\ref{fig: photon bins}d displays the single-photon detection timing histogram on an avalanche photodetector over four time-bins with a spacing of $\tau = 5680$~ns between pulsed excitations.

\begin{figure*}[htb!]%
\centering
\includegraphics[width=0.8\textwidth]{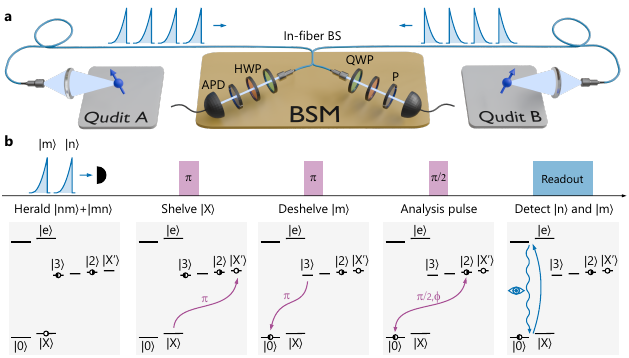}
\caption{
\textbf{Two-qudit network.} \textbf{(a)} Schematic of the experimental setup with two vacuum chambers labeled qudit A and B, each containing a single trapped \Ba ion.
Each atom is initialized in a qudit superposition state and emits a time-bin-encoded single photon entangled with its internal electronic state.
The two photons interfere at a beamsplitter and are detected by avalanche photodetectors (APDs).
Quarter- and half-waveplates (QWPs and HWPs) are used to compensate polarization rotation in the optical fiber.
Coincident detection of two photons in time-bins $n$ and $m$ ideally herald entangled atom-atom states of the form $\ket{n}_A\ket{m}_B \pm \ket{m}_A\ket{n}_B$, where $n,m \in \{0,1,2,3\}$ and $m > n$.
\textbf{(b)} Entanglement generation protocol. 
Following atomic excitation and successful heralding of two photons shown in (a) detected in time-bins $\ket{n}$ and $\ket{m}$, each atomic qudit is measured. 
First, residual population in the state $\ket{X}$ is shelved to the metastable $\ket{X'}$ state, then a series of shelving pulses sequentially transfers the expected qudit state to the $\ket{0}$ state for fluorescence detection. 
If there is no fluorescence in either state $\ket{n}$ or $\ket{m}$ for either atom, the event is discarded. Example is shown for entangled states involving $n=2$ and $m=3$.
}
\label{fig: ion-ion_setup}
\end{figure*}

We collect the emitted photon with a high-numerical-aperture objective and couple into a single-mode fiber.
Conditioned on successful photon collection, the atom-photon system is ideally left in a maximally entangled state
\begin{equation} \label{eqn:ionphoton}
    \ket{\psi}_\textrm{atom-photon} = \frac{1}{\sqrt{d}} \sum^{d-1}_{j=0}\ket{j}^{(a)} \ket{j}^{(p)},
\end{equation}
where $\ket{j}^{(p)}$ denotes a photon present in the $j^\text{th}$ time-bin at the fiber's input and $\ket{j}^{(a)}$ refers to the atom's internal qudit state.
For clarity, we omit the increment of atom state from $\ket{j}$ to $\ket{j+1}$ due to population swapping during time-bin generation.
These HD single photons have potential applications in quantum communication protocols and can, in principle, mediate HD entanglement with a second, spatially-separated qudit memory, establishing the foundation for networking multiple qudit-based QPUs, as depicted in Fig. \ref{fig: photon bins}a.

In the experiment, we generate two such atom-photon qudit entangled states from the two separate systems labeled A and B. The end-to-end photon collection and detection probabilities for each system are $p_{A} = 0.005$ and $p_{B} = 0.007$. 
Each probability is a product of the solid angle of light collection, optical losses (including fiber coupling and beamsplitter loss), and the detector efficiency, as detailed in Appendix \ref{M: Rate}.
Based on downstream measurements of entanglement between two qudit atomic memories as described below, we infer a bound \cite{Plenio2006} on the atom-photon entanglement fidelities of each atom-photon system to be $F>\{0.959, 0.931, 0.913\}$ for qudit dimensions $d=\{2,3,4\}$.


\section*{Networking qudit memories}

When two separate atom-photon HD entangled states are generated as outlined above, the atomic qudits can be entangled through Bell-state measurements (BSM) of the two HD photons~\cite{review_Bouchard2021}.
BSM is a cornerstone of quantum networking and essential for quantum repeater operation and establishing long-distance quantum communication links~\cite{review_Bouchard2021, review_Azuma2023}.
These measurements rely on two-photon interference, which erases which-path information and projects remote memories into entangled states.

\begin{figure*}[htb!]%
\centering
\includegraphics[width=\textwidth]{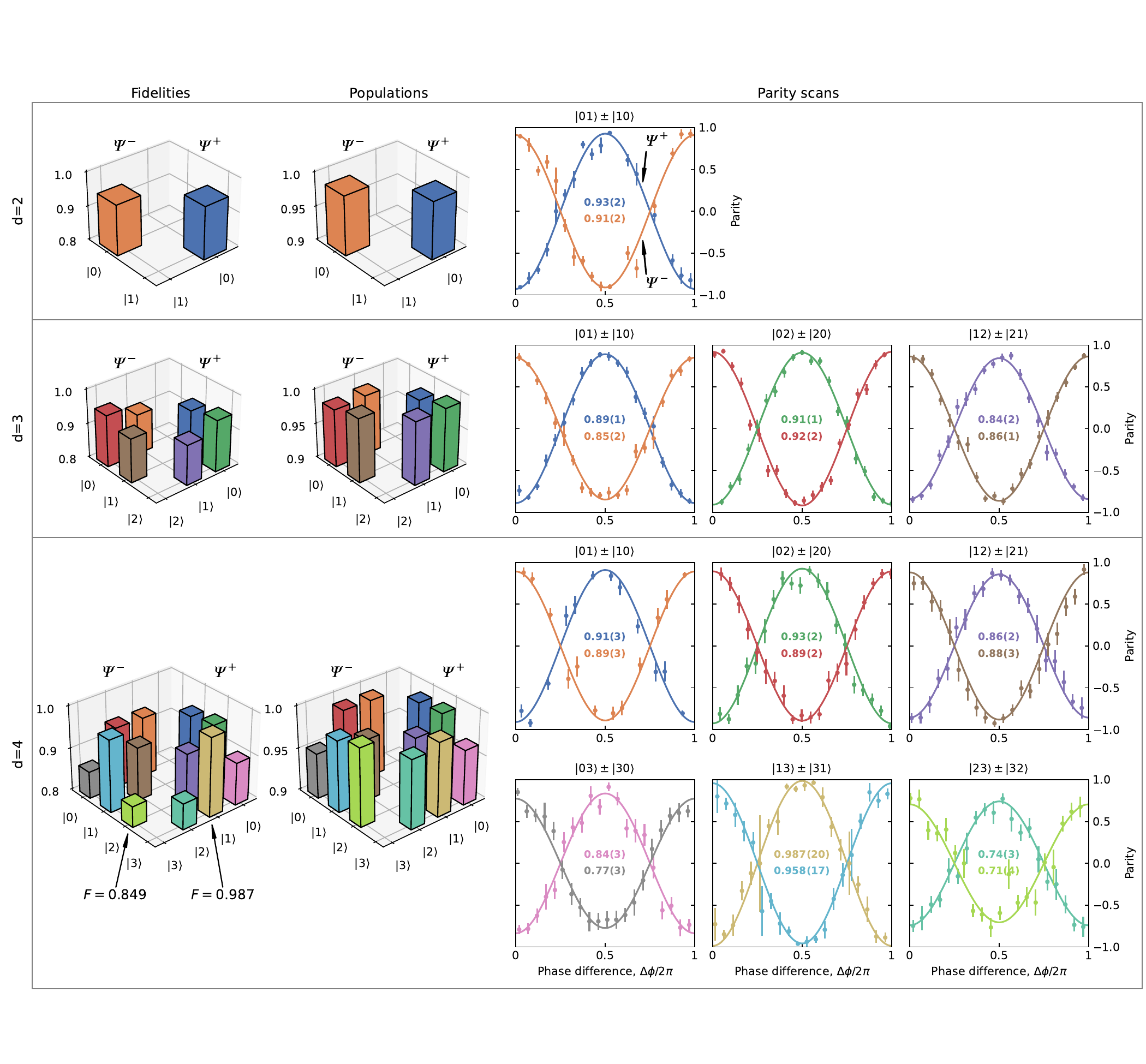}
\caption{\textbf{Qudit entangled state characterization.} 
The measured fidelity of each entangled qudit state ${\Psi^{\pm}=\ket{n}_A\ket{m}_B \pm \ket{m}_A\ket{n}_B}$ for dimensions $d=2$, $d=3$, and $d=4$ is displayed on the left column for each value of $n$ and $m$. 
The fidelities are calculated from the population measurements (middle column) and the contrast of the parity scans (right column).
The parity oscillations are measured by inserting global $\pi/2$ rotations with phase difference $\Delta\phi$ after heralding but before measurement, with statistical error bars and best-fit contrasts indicated in each plot. All data includes state preparation and measurement errors.}
\label{fig: ion-ion_data}
\end{figure*}

In the experiment, we entangle two qudit memories of dimensions $d=\{2,3,4\}$ residing in separate ion trap systems positioned $\sim 2$~m apart.
Fig.~\ref{fig: ion-ion_setup} shows the experimental apparatus, the entanglement generation protocol, and the corresponding control sequence (see details in Appendix \ref{M: Entangling protocol}).
We follow the same procedure as in the previous section, now exciting the two atomic qudits with synchronized short laser pulses and generating single photons from each atom. The photons are then spatially mode-matched on a beam splitter and detected on the output ports.
Detection of two photons in time-bins $n$ and $m$ ($m>n$) heralds the creation of a maximally entangled Bell state between the two memories, ideally given by
\begin{align} \label{eq: entangled states}
    \ket{\Psi^{\pm}} = \frac{\ket{n}_A \ket{m}_B \pm e^{i\phi}\ket{m}_A \ket{n}_B}{\sqrt{2}}.
\end{align}
The sign of the entangled state above is ``$+$" when the photons are detected on the same side of the beamsplitter and ``$-$" when they are detected on opposite sides. 
Importantly, the optical phases acquired when propagating from
the atoms to the beamsplitter cancel in the above state, as the optical path is stable between time-bins $n$ and $m$. 
The residual phase $\phi \ll 1$ in Eq. \ref{eq: entangled states} accounts for known relative level splittings between the two qudits, described below and in Appendix \ref{M: Rephasing}.

Once a BSM is performed and two detection events are registered in time-bins $n$ and $m$, the atomic qudit states are measured via standard fluorescence detection after coherently transferring populations from the $\ket{n}$ and $\ket{m}$ states of each atom to their $\ket{^2 S_{1/2}}$ ground state manifolds \cite{Senko2023, OReilly2024}.
The atomic qudit state detection fidelities are better than $0.99$, including imperfections in the population transfers. 

The fidelity of the entangled state with respect to an ideal Bell state is given by $ F = (P+C)/2$, where $P$ is the probability of occupying states $\ket{n}_A\ket{m}_B$ or $\ket{m}_A\ket{n}_B$ following a measurement, and $C/2$ is the magnitude of the off-diagonal coherence between the two states~\cite{Sackett2000}.
We measure the coherence by applying $\pi/2$ rotations with relative phase $\Delta\phi$ to the atoms before fluorescence detection. 
The value of $C$ is the contrast of the parity oscillations of the two atomic states as $\Delta\phi$ is scanned, and the states with opposite signs in Eq. \ref{eq: entangled states} are shifted by $\pi$ phase.
The measured populations and parity scans for all entangled states from Eq. \ref{eq: entangled states} are displayed in Fig.~\ref{fig: ion-ion_data}. 
We observe entangled state fidelities in a range between $0.849(23)$ and $0.987(13)$ (not adjusted for state preparation and measurement errors).

We now discuss the sources of observed fidelity imperfections.
Differential magnetic field drift between the two qudit systems will affect the phase of the entangled states in Eq. \ref{eq: entangled states}.
We mitigate this phase drift error under the assumption that the magnetic field drift varies over a time scale much slower than the entanglement generation rate. 
This allows us to monitor the phase for each heralded state in Eq. \ref{eq: entangled states} and keep track of its slow drift given the known sensitivity of each qudit state to magnetic fields. 
During data analysis, we feed-forward the phase accumulation to recover contrast degradation from slow magnetic field noise, as described in Appendix~\ref{M: Rephasing}. 

When the photons are produced, undesired decay events to the $\ket{X}$ level (Fig. \ref{fig: photon bins}c) are largely suppressed through polarization filtering. Residual events are converted into erasure errors by shelving the $\ket{X}$ state to the metastable $\ket{X'}$ state (Fig. \ref{fig: ion-ion_setup}b), thereby rendering the atom dark under fluorescence detection, as detailed in Appendix~\ref{M: Error budget}.
If either atom appears dark during state detection, the trial is discarded.
We have rejected between $6\%$ to $11\%$ of data depending on the qudit dimensionality allowing for similar fidelity improvements~\cite{high_F_Saha2024}.

Increased dimension requires more population swapping operations, increasing overall error for certain entangled Bell states.
The observed errors are primarily due to decoherence from fast differential magnetic field fluctuations and state swapping imperfections.
The measured fidelities of the various Bell states differ because of their unique sensitivities to magnetic field noise and contributions from state swapping errors, as detailed in the error budget in Appendix~\ref{M: Error budget}.
Some of these errors can be mitigated by using hyperfine states, which are insensitive to magnetic field drifts to the first order, and stabilizing the power and pointing noise of the population-swapping laser.
Nonetheless, several Bell states exhibit fidelities consistent with prior two-qubit entanglement benchmarks~\cite{high_F_Saha2024}.

\section*{Qudit Bell state success fraction}

\begin{figure}[tb!]%
\centering
\includegraphics[width=0.3\textwidth]{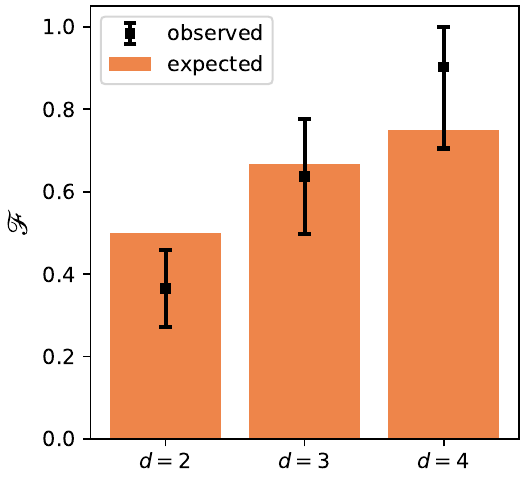}
\caption{
Measured Bell-state measurement (BSM) success fractions for qudit dimensions $d = \{2, 3, 4\}$ (black points and statistical error bars).
The success fraction increases with dimension according to the theoretical scaling $\mathscr{F} = 1 - 1/d$ (orange bars), which reflects the fraction of distinguishable antisymmetric Bell states.
}
\label{fig: ent_fraction}
\end{figure}

With every attempt, the probability of successfully generating atom-atom entanglement is $P_\text{ent} = \mathscr{F} p_{\text{A}} p_{\text{B}}$, where $\mathscr{F}$ is the qudit heralding success fraction.
The success probability of Bell-state measurements for two qubits ($d=2$) using passive linear optics is fundamentally limited to $\mathscr{F} = 1/2$, because two out of the four Bell states produce indistinguishable coincidence detection patterns~\cite{Calsamiglia2001, Ghosh2001}.
This limitation can be overcome by using additional photonic qubits~\cite{ancil_Grice2011, ancil_Bayerbach2023}, entanglement across multiple degrees of freedom~\cite{hyper_Kwiat1998, hyper_Schuck2006, hyper_Barbieri2007, hyper_Wang2016, hyper_Li2017}, nonlinear interactions at the single-photon level~\cite{NL_Kim2001}, or complex feed-forward control~\cite{adaptive_Knill2001, adaptive_Rad2025, HD_photons_Bartolucci2023}.
Alternatively, HD single photons enable higher BSM success probabilities without additional complexity.

For two qudits each prepared in an equal superposition of its $d$ states and correlated with its photon qudit as in Eq. \ref{eqn:ionphoton}, there are $d^2$ possible input states to the beamsplitter.  
Of these, $d$ states produce photons in the same time-bin ($n=m$), which do not produce entanglement and are discarded.
After heralding and detection, this leaves $d^2-d$ entangled states of the qudits in the form of Eq.~\ref{eq: entangled states}. 
The qudit entanglement success fraction is therefore $\mathscr{F} = 1-1/d$~\cite{review_Bouchard2021, HD_ent_rate_Yamazaki2024}.

Fig.~\ref{fig: ent_fraction} displays the observed and ideal qudit heralding success probabilities for dimensions $d = \{2,3,4\}$.
The success fractions $\mathscr{F}$ were extracted from the experimental atom-atom entanglement success probabilities $P_\text{ent}$ by factoring out the known efficiencies $p_\text{A}$ and $p_\text{B}$ for each atom-photon system (see Appendix~\ref{M: Rate}).
The observed success fraction increases with dimension as expected,
overcoming the fundamental Bell-state detection limit.

\section*{Outlook}

Qudit memories and higher-dimensional photons can significantly expand the capacity of quantum information networks, and offer key advantages for advancing both quantum computing and quantum communication.
While the form of entanglement generated in this work is limited to Bell states stored in qudits, this can be straightforwardly expanded to HD entanglement spanning every level of the qudit manifold. 
For example, the time-bin qudit photons can be converted to multiple paths that allow programmable all-to-all interference, which can herald arbitrary qudit memory entangled states~\cite{HD_photons_Erhard2018}.
Alternatively, HD entanglement between two qudit memories can be established via a single HD photon using pitch-and-catch schemes: one memory emits the photon, which is then absorbed by a second memory, directly transferring the quantum state~\cite{atom_photon_Cirac1997, atom_photon_Wilk2007, atom_photon_Kurz2014}.
Distributing such HD entanglement across remote quantum memories offers a resource-efficient pathway for quantum-information transfer and may thus play an important role in scalable quantum network architectures.

\textbf{Acknowledgments.}
This work is supported by the DOE Quantum Systems Accelerator (DE-FOA-0002253) and the NSF STAQ Program (PHY-1818914). A.K. is supported by the AFOSR National Defense Science and Engineering Graduate (NDSEG) Fellowship.

\bibliographystyle{apsrev4-1} 
\bibliography{bibliography}

\clearpage

\begin{widetext}
\section*{Appendices}\label{sec: Methods}

\subsection{Remote entanglement rate and success probability}\label{M: Rate}

Fig.~\ref{fig: M:rate} presents the remote qudit-qudit entanglement generation rates and entanglement generation success probabilities measured across multiple data sets for each qudit with dimension $d$.
Each data point corresponds to a separate acquisition run.
Dashed lines indicate the average rate for each value of $d$.

The remote entanglement rate is written as
\begin{equation}
    R = \frac{\mathscr{F} p_A p_B}{\tau_0 + \tau_\textrm{bin} (d-1)},
\end{equation}
where
$\tau_0$ is the entanglement attempt time excluding time-bin generation and
$\tau_\textrm{bin}$ is the overhead for the generation of extra time-bins. 
$\mathscr{F} = 1 - 1/d$ is the dimension-dependent entanglement success fraction.
The overhead for adding extra dimensions to qudits may outweigh the gains in the entanglement rate, so the optimal trade-off may depend upon how quickly the extra dimensions (in our case time-bins) can be generated.

\begin{figure*}[h!]%
\centering
\includegraphics[width=0.8\textwidth]{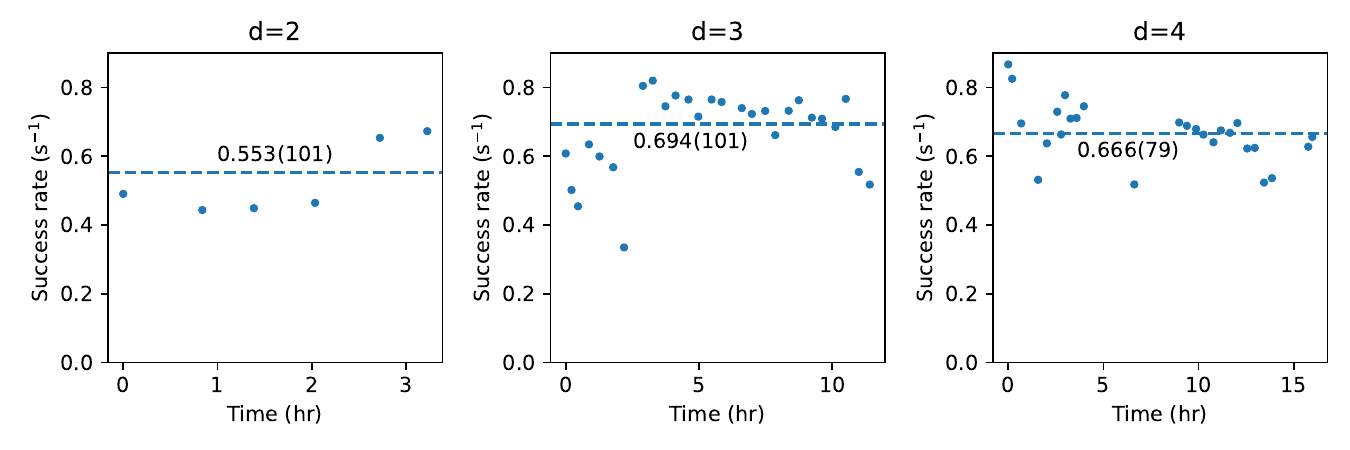}
\includegraphics[width=0.8\textwidth]{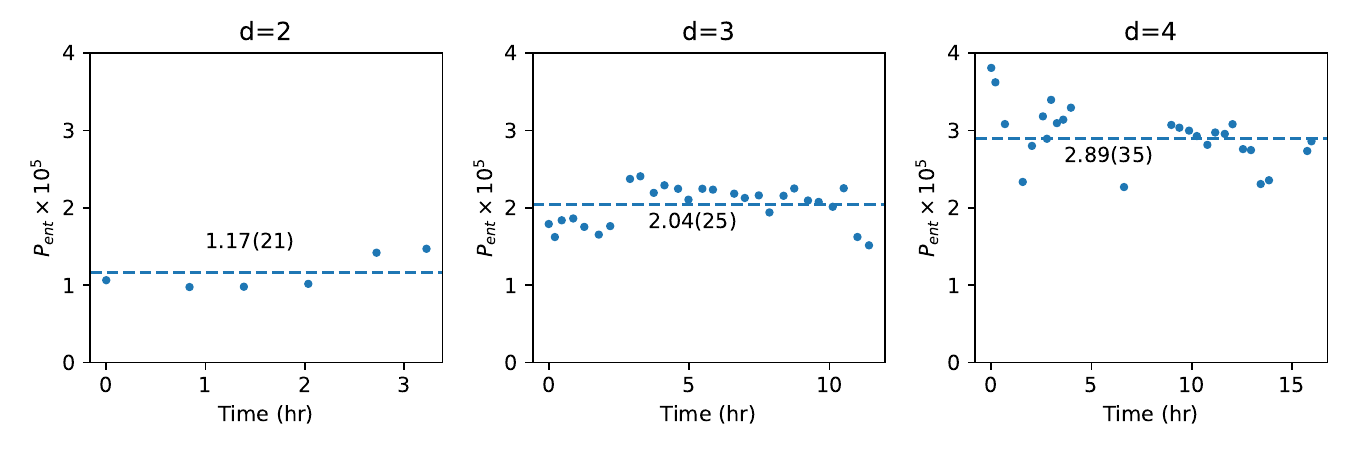}
\caption{Remote entanglement rates and total entanglement success probabilities for dimensions $d = \{2, 3, 4\}$.
Each point represents a separate data set collected over time.
Dashed lines indicate the average entanglement rate for each qudit dimension.
}
\label{fig: M:rate}
\end{figure*}

The entanglement rate is primarily limited by the duration of single-qubit operations, which depends on the available optical power and the optics used for beam focusing.
A further limitation arises from the need to synchronize time-bin generation with the ion's motional period.
This constraint may be relaxed by increasing trap frequencies 
so that the time-bin delay is much shorter than the period of atomic motion~\cite{li2025}, or by cooling ions to near the motional ground state, perhaps through continuous sympathetic cooling~\cite{OReilly2024}.

The entanglement rates vary with qudit dimension due to the increasing overhead required to generate and manipulate additional time-bins.
For $d = \{2, 3, 4\}$, the entanglement cycle periods are $\tau_\textrm{cycle} \approx \{17.0, 22.7, 34.1\}~\mu\textrm{s}$, respectively.
These durations also depend on the Rabi frequency of the $1762$~nm single-qubit rotations, which are regularly re-calibrated during data collection to ensure high-fidelity rotations.
Table~\ref{tab:additional_data} lists other relevant metrics from our experiment for each qudit dimension, including the number of successes and attempts.

\begin{table}[ht]
\centering
\renewcommand{\arraystretch}{1.2}
\begin{tabular}{|c|c|c|r|r|r|c|c|}
\hline
\textit{d} & 
\begin{tabular}[c]{@{}c@{}}Attempt\\length\\(µs)\end{tabular} & 
\begin{tabular}[c]{@{}c@{}}Attempt\\rate\\(kHz)\end{tabular} & 
\multicolumn{1}{c|}{\text{Time (s)}} & 
\multicolumn{1}{c|}{\text{Successes}} & 
\multicolumn{1}{c|}{\text{Attempts}} & 
\begin{tabular}[c]{@{}c@{}}Experimental\\entanglement\\rate (s$^{-1}$)\end{tabular} & 
\begin{tabular}[c]{@{}c@{}}$P_\textrm{ent}$\\($\times 10^{-5}$)\end{tabular} \\
\hline
2 & 17   & 58  & 9,844   & 5,121   & 437,972,714   & 0.553 & 1.17 \\
3 & 22.7 & 44  & 35,342  & 23,427  & 1,149,607,096 & 0.694 & 2.04 \\
4 & 34.1 & 29  & 31,315  & 19,941  & 689,981,646  & 0.666 & 2.89 \\
\hline
\end{tabular}
\caption{Experimental metrics for different qudit dimensions $d=\{2, 3, 4\}$. Attempt length increases with dimension due to the need for more swaps between levels. Wall-clock time and experimental entanglement rate are calculated with experimental overhead included. The entanglement generation success probability $P_\textrm{ent}$ is computed from the number of successes and attempts.}
\label{tab:additional_data}
\end{table}

The atom-atom entanglement success probability is given by $P_{ent} = \mathscr{F} p_A p_B$, as defined in the main text. The reported values of $P_\text{ent}$ are directly calculated from the total number of successes and attempts during our experiment. 
The probability of light collection and detection from each chamber denoted $p_{A}$ and $p_{B}$ are products of various efficiencies and losses as listed in Table~\ref{tab:light_coll_prob}. 
To extract the fractions $(\mathscr{F})$ as shown in Fig.~\ref{fig: ent_fraction}d, we factor $p_A$ and $p_B$ from the directly measured atom-atom entanglement success probability $(P_\text{ent})$. For $d=\{2, 3, 4\}$, the extracted success fractions are $\mathscr{F} = \{0.37 (11), 0.64 (17), 0.90 (24) \}$ respectively, as shown in Fig.~\ref{fig: ent_fraction}.

\begin{table}[h!]
\centering
\begin{tabular}{|l | c | c|}
\hline
Component              &      System A &  System B \\
\hline
Lens solid angle       &     0.1      & 0.2   \\
Fiber coupling         &     0.30(4)  & 0.20(3)   \\
Trap clipping          &     0.78(2) &  0.97(1)   \\
Optical losses         &    0.90(2)  &  0.80(2)   \\
APD quantum efficiency &    0.65(3)   & 0.65(3)   \\
Excitation probability &     0.70(5)    &0.70(5)   \\
Branching ratio        &     0.486     & 0.486   \\\hline
Total                  &$p_A=0.0047(8)$& $p_B=0.0069(12)$ \\
\hline
\end{tabular}
\label{tab:light_coll_prob}
\caption{Breakdown of photon collection and detection efficiencies $p_A$ and $p_B$ for individual atom-photon systems A and B.}
\end{table}

\subsection{The entangling protocol}\label{M: Entangling protocol}

The ions are first Doppler-cooled and optically pumped into the ground state $\ket{0}$.
Each ion is then prepared in an equal superposition so the resulting two-ion state can be written as:
\begin{equation}
    \ket{\psi_0} = \frac{1}{d} \sum^{d-1}_{j=0}\ket{j}^{(a)}_A \otimes \sum^{d-1}_{q=0}\ket{q}^{(a)}_B
\end{equation}
where $d$ is the qudit dimension.
The superscript $(a)$ denotes an atomic ion state, and the labels $\{A, B\}$ refer to the ions in the A and B chambers, respectively.

To generate a photon in a specific time-bin, the ion population in $\ket{0}$ is promoted to the excited state $\ket{e}$, from which it may spontaneously decay back to $\ket{0}$, emitting a single photon in time-bin $j$. 
The operator for such a process is:

\begin{equation}
    A_j = \ket{j}^{(a)}\bra{j}^{(a)} \left(\sqrt{p} e^{i \mathbf{\Delta k} \cdot \mathbf{r}_{A,B}(t_j)} c_j^\dagger + \sqrt{1-p}\right) + \sum_{q \neq j}^{d-1} \ket{q}^{(a)} \bra{q}^{(a)}
\end{equation}
where $p$ is the overall probability to detect the photon,
$c_j^\dagger$ is the photon creation operator for time-bin $j$,
${\mathbf{\Delta k}=\mathbf{k}'-\mathbf{k}}$ is the difference between the excitation pulse wavevector $\mathbf{k}'$ and the emitted photon wavevector $\mathbf{k}$;
and $\mathbf{r}_{A,B} (t_j)$ is the atomic position operator for ion A or B.

The two ion-photon joint states at the beamsplitter input, conditioned on successful photon emission in time-bin $n$ and collection into an optical fiber, is given by:

\begin{align}
    \ket{\psi_1} = \ket{n}^{(a)}_A \sum^{d-1}_{j>n}\ket{j}^{(a)}_B \ket{n}^{(p)}_A e^{i \phi_{A,n}} + \ket{n}^{(a)}_B \sum^{d-1}_{j>n}\ket{j}^{(a)}_A \ket{n}^{(p)}_B e^{i \phi_{B,n}}
\end{align}
where
$\phi_{\{A,B\},n} = \mathbf{\Delta k}\cdot\mathbf{r}_{\{A,B\}}(t_n) + \phi^\textrm{opt}_{\{A,B\}}$ accounts for the optical and motional phases;
$\mathbf{r}_{\{A,B\}}(t_e) $ is the position of an ion $\{A,B\}$ at the time $t_n$; 
the superscript $(p)$ denotes photonic modes, and ${\{A,B\}}$ subscripts indicates the presence of a photon in the input modes $A$ or $B$ of a beamsplitter.

The photons are interfered on a beamsplitter, so we apply the beamsplitter transformation $\ket{n}_{\{A,B\}}\rightarrow\ket{n}_C \pm \ket{n}_D $, yielding the atom-atom state conditioned on a detection in output modes $C$ or $D$:

\begin{align}
    \ket{\psi_2} = &\ket{n}^{(p)}_C \left( \ket{n}^{(a)}_A \sum^{d-1}_{j>n}\ket{j}^{(a)}_B e^{i \phi_{A,n}}
                    + \ket{n}^{(a)}_B \sum^{d-1}_{j>n}\ket{j}^{(a)}_A e^{i \phi_{B,n}}\right) \nonumber \\
                    +&\ket{n}^{(p)}_D \left( \ket{n}^{(a)}_A \sum^{d-1}_{j>n}\ket{j}^{(a)}_B e^{i \phi_{A,n}}
                    - \ket{n}^{(a)}_B \sum^{d-1}_{j>n}\ket{j}^{(a)}_A e^{i \phi_{B,n}}\right).
\end{align}

To generate the other time-bins, we swap the population between $\ket{0}$ and $\ket{j}$ and apply an excitation pulse.
Upon successful detection of a second photon in time-bin $m > n$, the entangled atom-atom state becomes:

\begin{align}
    \ket{\psi_3} = &\left(\ket{n}^{(p)}_C \ket{m}^{(p)}_C - \ket{n}^{(p)}_D \ket{m}^{(p)}_D \right) 
                    \left(\ket{n}_A^{(a)}\ket{m}_B^{(a)} e^{i \Delta \phi} + \ket{m}_A^{(a)}\ket{n}_B^{(a)} \right) \nonumber \\
                  +&\left(\ket{n}^{(p)}_C \ket{m}^{(p)}_D - \ket{n}^{(p)}_D \ket{m}^{(p)}_C \right) 
                    \left(\ket{n}_A^{(a)}\ket{m}_B^{(a)} e^{i \Delta \phi} - \ket{m}_A^{(a)}\ket{n}_B^{(a)} \right)
\end{align}
where
$\Delta \phi= \left( \phi_{A,n} -\phi_{A,m} \right) - \left( \phi_{B,n} -\phi_{B,m} \right)$ is differential phase for time-bins $n$ and $m$.

In our implementation, we assume synchronization between pulsed excitations and ion motion periods, and negligible optical phase drift between time-bin generations.
Under this assumption, optical and motional phase contributions cancel for time-bins $n$ and $m$, reducing the overall phase to a global phase that can be neglected.

Note that in our experiments, the final atomic states are different due to the swap pulses used for time-bin generation.
This advances the state as $\{n,m\} \rightarrow \{n+1,m+1\}$ (if $m=d$, then $m\rightarrow0$).

\subsection{Mitigation of slow magnetic field drifts through feed-forward}\label{M: Rephasing}

The entangled states of Eq, \ref{eq: entangled states} are insensitive to common-mode magnetic field noise due to their symmetry. However, differential magnetic field noise between the qudits will affect the phase of their coherence relative to an outside local oscillator (or other quantum systems to be networked). We can absorb that phase by writing the effective entangled state as
\begin{equation} \label{eq: EntPhase}
\ket{\psi(t')} = \frac{\ket{n}_A \ket{m}_B \pm e^{-i \phi} \ket{m}_A \ket{n}_B}{\sqrt{2}}.
\end{equation}
Here $\phi = \delta\!B \gamma_{nm}T$, where $\delta\!B$ is the average magnetic field difference between the two qudit positions A and B over time interval $T$ between the heralding detection and the final analysis $\pi/2$ pulse.
The magnetic field sensitivity $\gamma_{nm}$ depends on the relative Zeeman shifts of the particular atomic levels $\ket{n}$ and $\ket{m}$ involved, given by their relative magnetic g-factors. In addition, the atoms are shelved and swapped through various field-sensitive states depending on the dimension $d$ of the final entangled state, resulting in a complicated overall sensitivity of each entangled state in Eq.~\ref{eq: EntPhase} to the magnetic field. 
Table~\ref{tab: Magnetic field sensitivity} summarizes the effective sensitivities $\gamma_{nm}$ to magnetic field for all measured Bell states in dimensions $d = \{2,3,4\}$.

During data analysis, we make fine corrections to the phase of each coherence measurement in the parity scan through a feed-forward technique that effectively compensates for a slowly-varying differential magnetic field. 
We determine the particular differential magnetic field $\delta\!B$ at each time that minimizes the deviation of each adjusted phase from that expected by the known phase sensitivity from Table~\ref{tab: Magnetic field sensitivity}.
This feed-forward adjustment effectively rephases the entangled states and improves the measured state coherence (parity scan contrast) by up to $\approx22\%$ for the most sensitive state.
\begin{table}[h!]
\centering
\begin{tabular}{|c|r|r|r|}
\hline
 & \multicolumn{3}{c|}{Phase sensitivity (rads/mG), $\gamma_{nm}$} \\
\hline
{Heralded Bell state} & d=4 & d=3 & d=2 \\ 
\hline
$\ket{10} \pm \ket{01}$ & 0.31   & \hspace{1cm} 0.61   & 0.49 \\
$\ket{20} \pm \ket{02}$ & -1.24  & \hspace{1cm} -0.92 & -- \\
$\ket{30} \pm \ket{03}$ & 1.71  & \hspace{1cm} --     & -- \\
$\ket{12} \pm \ket{21}$ & 1.64  & \hspace{1cm} 1.65  & -- \\ 
$\ket{13} \pm \ket{31}$ & -0.50 & \hspace{1cm} --     & -- \\ 
$\ket{23} \pm \ket{32}$ & -3.22 & \hspace{1cm} --     & -- \\
\hline
\end{tabular}
\caption{ Bell state phase sensitivities to magnetic field for all heralded states and qudit dimension $d=\{2,3,4\}$. These are given by the magnetic g-factors of the relevant states along with dwell times for shelving and swapping operations for each protocol. 
}
\label{tab: Magnetic field sensitivity}
\end{table}

Figure~\ref{fig: Phase fit} shows the measured phase evolution over hours of time for each Bell state, with a phase shift of $\pi$ added manually to all odd parity states $\ket{\Psi^-}$ to bring the curves together. 
The drifting phase of the coherence for each state in Eq.~\ref{eq: EntPhase} was fitted to a uniform magnetic field that best reproduced the known ratio of phase shifts given by Table \ref{tab: Magnetic field sensitivity}.  

\begin{figure*}[h!]%
\centering
\includegraphics[width=\textwidth]{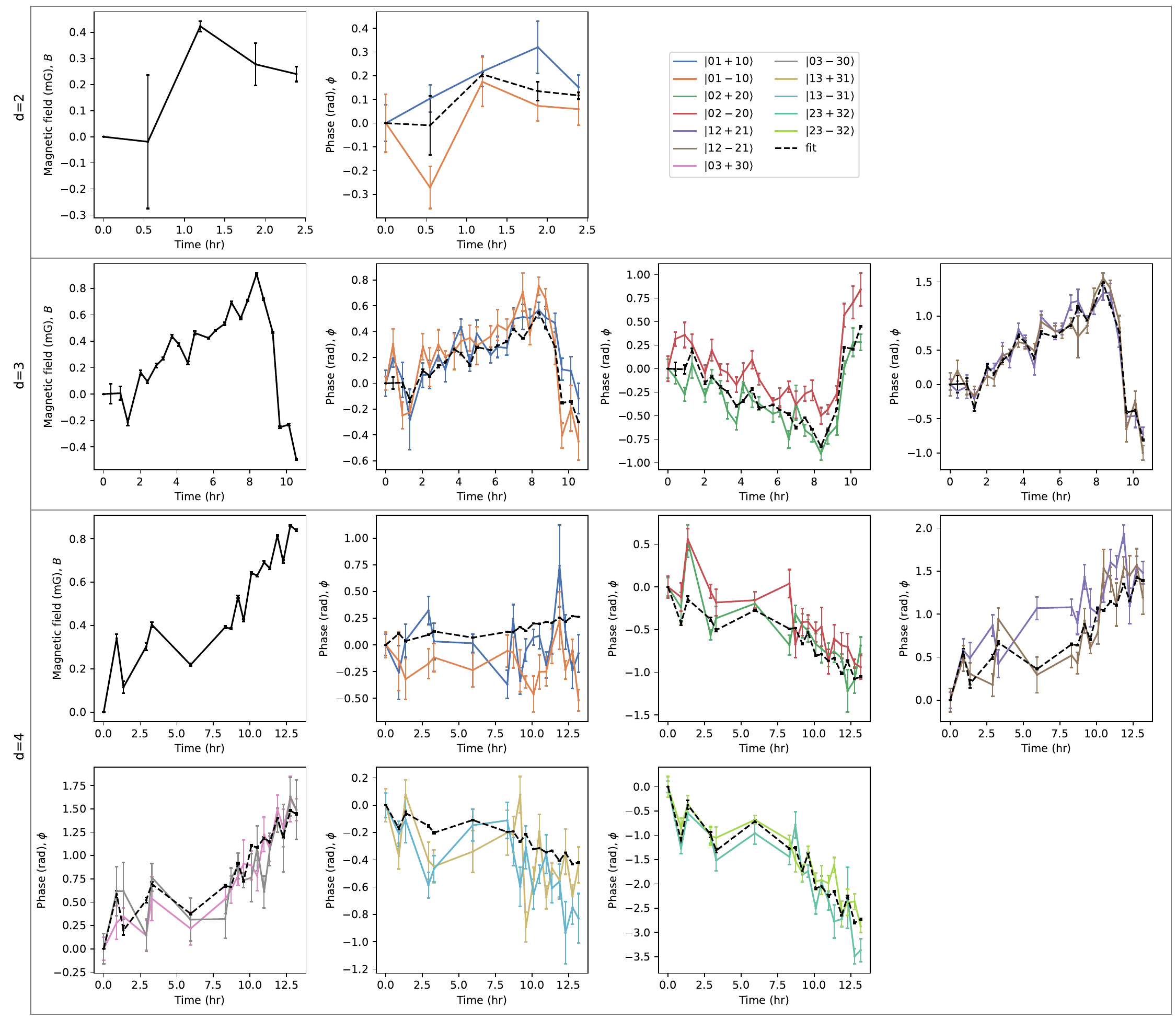}
\caption{
Differential magnetic field drift and phase feed-forward. 
Solid points are measured phase drifts of all entangled state coherences for qudit dimensions $d=2,3,4$. 
The dashed lines are fits of the state phases at each point in time to a particular differential magnetic field $\delta\! B$ between the two systems that minimizes the deviation from the phase shift $\phi = \delta\!B \gamma_{nm}(d) T$ using the state sensitivities from Table \ref{tab: Magnetic field sensitivity}. 
The fitted differential magnetic field is shown in the first plot of each family of plots.
}
\label{fig: Phase fit}
\end{figure*}

The rephasing procedure improves fidelity estimates by compensating for slow environmental drifts and ensures accurate characterization of the entangled states.
All results shown in the main text include this phase correction.
We show the measured populations, contrast and fidelity for each entangled state in Table~\ref{tab:fid_pop_con}. 
We also show the detection probabilities for each ion-ion pair at the points maximum contrast in Figure~\ref{fig: ion-ion_data analysis}.
Note, that the values shown in Figures~\ref{fig: ion-ion_data populations} and ~\ref{fig: ion-ion_data analysis} are calculated assuming the Central Limit Theorem, unlike the results shown in Table~\ref{tab:fid_pop_con}.

\subsection{Error budget}\label{M: Error budget}

Table~\ref{tab: Error budget} summarizes the estimated errors contributing to the infidelity of the final entangled states.
The total error is separated into two categories: errors common to all entangled states, and those specific to individual Bell states.
The common error sources contribute approximately $1.51\%$ to the total infidelity across all measurements.
$0.5\%$ of the common error sources are due to state preparation and measurement (SPAM), which is limited by our 1762 nm shelving fidelity to the detection dark state. 
Our dark state discrimination error is negligible.

\begin{table}[!htb] \label{tab: Error budget}
\centering
\renewcommand{\arraystretch}{1.2}
\noindent

\textbf{Common errors ($\%$)}
\vspace{2mm}

\centering
\begin{tabular}{|l|c|} \hline
 SPAM                                   & 0.5 \\ \hline
 Photon wavepacket overlap              & 0.2 \\ \hline
 Atom recoil, $\delta t = 50$~ns        & 0.3 \\  \hline
 Background counts                      & {$<0.2$}\\   \hline
 Atom recoil, $\omega_{qi}$ fluctuation & {$<0.1$} \\  \hline
 Beamsplitter imperfection              & {$<0.1$}\\  \hline
 Residual erasure errors                & {$<0.1$} \\  \hline
 Micromotion                            & {$<0.01$}\\ \hline
 \textbf{TOTAL common errors} (\%)      & {$\mathbf{1.51}$}\\ \hline
\end{tabular}

\vspace{3mm}
\textbf{Errors for individual Bell states ($\%$)}
\vspace{2mm}

\begin{tabular}{|l|c|c|c|c|c|c|c|c|c|} \hline
 & \multicolumn{3}{c|}{$d=4$} & \multicolumn{3}{c|}{$d=3$} & \multicolumn{3}{c|}{$d=2$} \\ \cline{2-10}
\textbf{Bell state} & Decoherence & 1762 error & \textbf{TOTAL} (\%) & Decoherence & 1762 error & \textbf{TOTAL} (\%) & Decoherence & 1762 error & \textbf{TOTAL} (\%) \\ \hline
$\ket{01} \pm \ket{10}$ & 0.1 & 0.98 & \textbf{2.59} & 0.2 & 0.98 & \textbf{2.69} & 0.1 & 8.6 & \textbf{10.21} \\ \hline
$\ket{02} \pm \ket{20}$ & 0.9 & 2.61 & \textbf{5.02} & 0.5 & 10.76 & \textbf{12.77} & - & - & - \\ \hline
$\ket{12} \pm \ket{21}$ & 1.6 & 7.68 & \textbf{10.79} & 1.6 & 7.12 & \textbf{10.23} & - & - & - \\ \hline
$\ket{03} \pm \ket{30}$ & 1.7 & 7.1 & \textbf{10.31} & - & - & - & - & - & - \\ \hline
$\ket{13} \pm \ket{31}$ & 0.2 & 1.19 & \textbf{2.9} & - & - & - & - & - & - \\ \hline
$\ket{23} \pm \ket{32}$ & 5.9 & 6.23 & \textbf{13.64} & - & - & - & - & - & - \\ \hline
\end{tabular}

\caption{
Error budget.
}
\end{table}

State-specific errors vary depending on the magnetic field sensitivity and the number of single-qubit operations required for each of the final Bell states.
Decoherence significantly degrades the Bell states that are more sensitive to magnetic field.
In addition, we noticed a significant double-excitation error that arises when $m-n=1$ for photons detected in time-bins $m, n$.
This is caused by infidelity in our 1762 nm shelving pulses during the entanglement generation procedure, as any population leftover in the ground state from the swap pulses may be excited a second time. 
This error is suppressed for entangled states with $m-n>1$, because the subsequent swap pulses act as a shelving pulse for the ``leftover" population. Note that the final atomic states are different from the heralded time-bin detections because of the swap pulses, specifically the atomic state gets advanced by 1, $n \rightarrow n+1$. For example, the states $|01\rangle \pm |10\rangle$ are heralded by detections in time bins (0,3).

Bell states with minimal sensitivity to these effects (such as $\ket{01 \pm 10}$ or $\ket{13 \pm 31}$) exhibit fidelities consistent with our prior work on remote entanglement with time-bin photon encoding~\cite{high_F_Saha2024}.
In contrast, states like $\ket{23 \pm 32}$, require additional single qubit rotations and are more prone to decoherence due to higher sensitivities to magnetic field, exhibit significantly higher error rates.

In our protocol, we post-select on detecting both ions in $n$ or $m$ states via state-selective fluorescence, detecting them sequentially.
Any event in which an ion is detected as dark for both states is treated as an error and ignored.

Finally, we mitigate polarization errors caused by inhomogeneous birefringence of optical components (such as vacuum windows) that can mix $\sigma^+$ and $\pi$ polarization and cause false-positive heralding events.
These errors are accompanied by ion population decaying to the $\ket{X}$ state and are mitigated by driving this population to the $\ket{X'}$ level and rendering the ion dark during both our state detections (for $n$ and $m$ states), thus converting false heralds into erasure errors as described in Ref.~\cite{high_F_Saha2024}.
We reject $\{11.5,6.4,9.1\}\%$ of data for qudit dimensions $d=\{2,3,4\}$, correspondingly.
Discarding the cases outside the intend state subspace allows for similar improvements in state fidelities.


\begin{figure*}[h!]%
\centering
\includegraphics[width=\textwidth]{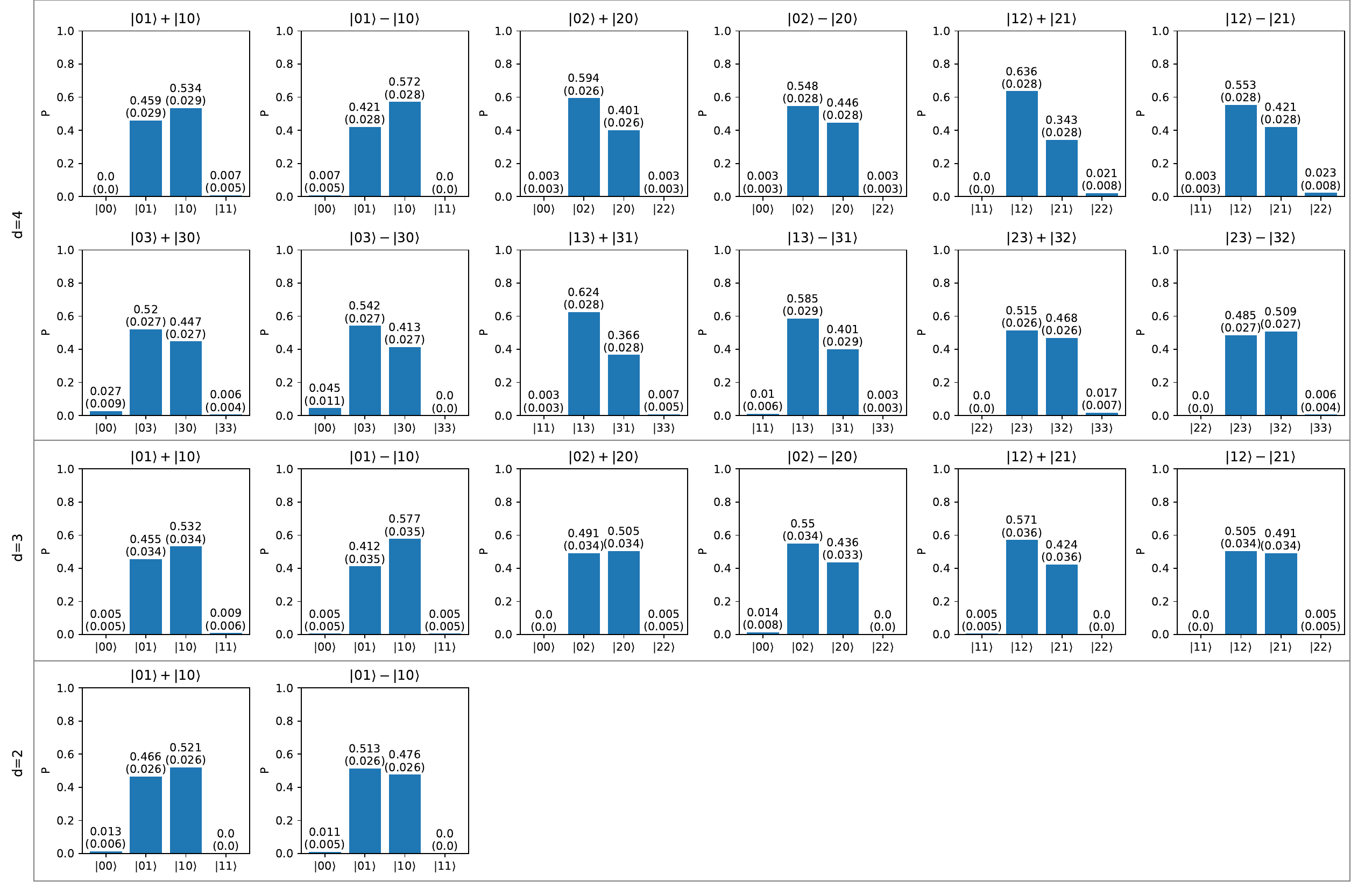}
\caption{
    The states of ions A and B ion after the entanglement heralding.
}
\label{fig: ion-ion_data populations}
\end{figure*}

\begin{figure*}[h!]%
\centering
\includegraphics[width=\textwidth]{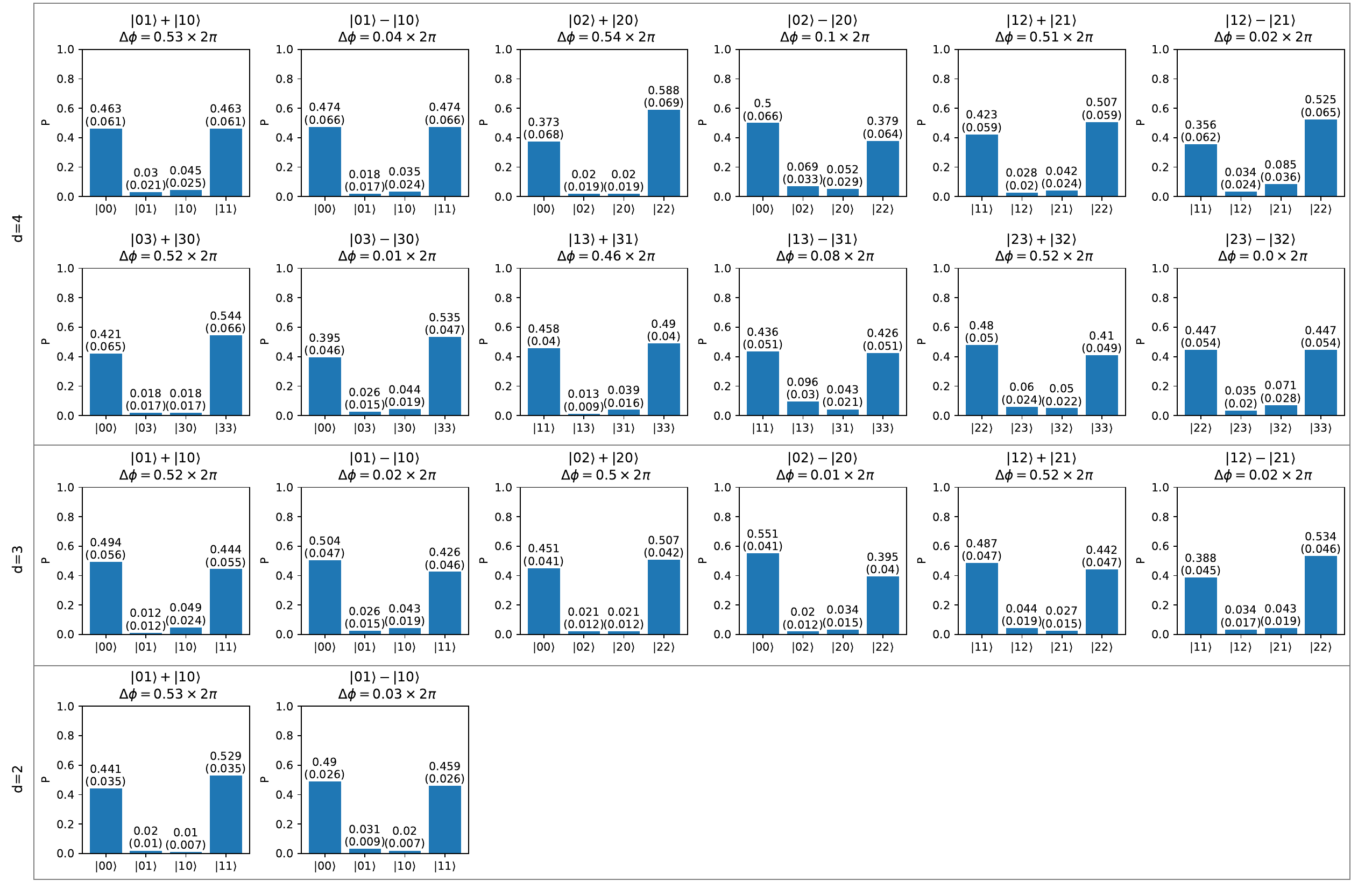}
\caption{
        The states of ions A and B ion after entanglement heralding followed by analysis $\pi/2$-rotation with the phase difference $\Delta\phi$.
}
\label{fig: ion-ion_data analysis}
\end{figure*}

\begin{table}[!htb]
\centering
\renewcommand{\arraystretch}{1.2}
\noindent

\textbf{Dimension $d=2$}
\vspace{2mm}

\centering
\begin{tabular}{|c|c|c|c|}
\hline
\textbf{State} & \textbf{Fidelity} & \textbf{Populations} & \textbf{Contrast} \\
\hline
$\ket{01}+\ket{10}$ & 0.950 (19) & 0.986 (6) & 0.909 (33) \\
$\ket{01}-\ket{10}$ & 0.941 (15) & 0.988 (5) & 0.891 (26) \\
\hline
\end{tabular}

\vspace{3mm}
\textbf{Dimension $d=3$}
\vspace{2mm}

\begin{tabular}{|c|c|c|c|}
\hline
\textbf{State} & \textbf{Fidelity} & \textbf{Populations} & \textbf{Contrast} \\
\hline
$\ket{01}+\ket{10}$ & 0.950 (19) & 0.984 (8) & 0.909 (33) \\
$\ket{01}-\ket{10}$ & 0.941 (15) & 0.987 (8) & 0.891 (26) \\
\hline
$\ket{02}+\ket{20}$ & 0.959 (13) & 0.993 (5) & 0.925 (23) \\
$\ket{02}-\ket{20}$ & 0.943 (11) & 0.984 (8) & 0.893 (17) \\
\hline
$\ket{12}+\ket{21}$ & 0.917 (13) & 0.992 (6) & 0.857 (17) \\
$\ket{12}-\ket{21}$ & 0.927 (18) & 0.993 (5) & 0.881 (26) \\
\hline
\end{tabular}

\vspace{3mm}
\textbf{Dimension $d=4$}
\vspace{2mm}

\begin{tabular}{|c|c|c|c|}
\hline
\textbf{State} & \textbf{Fidelity} & \textbf{Populations} & \textbf{Contrast} \\
\hline
$\ket{01}+\ket{10}$ & 0.950 (19) & 0.991 (5) & 0.909 (33) \\
$\ket{01}-\ket{10}$ & 0.941 (15) & 0.992 (5) & 0.891 (26) \\
\hline
$\ket{02}+\ket{20}$ & 0.959 (13) & 0.993 (4) & 0.925 (23) \\
$\ket{02}-\ket{20}$ & 0.943 (11) & 0.992 (5) & 0.893 (17) \\
\hline
$\ket{12}+\ket{21}$ & 0.917 (13) & 0.977 (9) & 0.857 (17) \\
$\ket{12}-\ket{21}$ & 0.927 (18) & 0.973 (9) & 0.881 (26) \\
\hline
$\ket{03}+\ket{30}$ & 0.901 (18) & 0.966 (10) & 0.836 (27) \\
$\ket{03}-\ket{30}$ & 0.864 (18) & 0.953 (11) & 0.774 (25) \\
\hline
$\ket{13}+\ket{31}$ & 0.987 (13) & 0.988 (6) & 0.987 (20) \\
$\ket{13}-\ket{31}$ & 0.971 (12) & 0.985 (7) & 0.958 (17) \\
\hline
$\ket{23}+\ket{32}$ & 0.862 (18) & 0.982 (7) & 0.741 (29) \\
$\ket{23}-\ket{32}$ & 0.849 (23) & 0.993 (4) & 0.706 (41) \\
\hline
\end{tabular}

\caption{
Measured fidelities, populations, and contrasts for each entangled Bell state for qudit dimensions $d=\{2,3,4\}$. 
}
\label{tab:fid_pop_con}
\end{table}

\end{widetext}

\end{document}